\documentclass[12pt,a4paper]{article}
\usepackage[utf8]{inputenc}
\usepackage{amssymb,amsmath}
\usepackage[margin=.8in]{geometry}
\usepackage{orcidlink}
\usepackage{authblk}

\DeclareMathOperator{\diag}{diag}
\DeclareMathOperator{\ord}{\mathcal O}
\DeclareMathOperator{\dif}{d\!}
\DeclareMathOperator{\e}{e}
\DeclareMathOperator{\re}{Re}
\DeclareMathOperator{\im}{Im}

\DeclareMathOperator{\id}{id}

\def\der{{\rm d}}
\def\hv#1{\hat \alpha^{#1}}
\def\sv{\sigma v}
\def\tsv{(\tilde{\sigma v})}
\def\Eg{E_\gamma}
\def\Mdm{m_\chi}

\def\amp{\mathcal A}

\def\full{\texttt{`full\_\,SE'}}
\def\fsr{\texttt{`FSR\_\,SE'}}
\def\fix{\texttt{`fixed\_\,noSE'}}
\def\endp{\texttt{`endp\_SE'}}

\def\neu#1{\chi^0_#1}
\def\cha#1{\chi^+_#1}
\def\acha#1{\chi^-_#1}
\def\chapm#1{\chi^\pm_#1}

\def\nn#1{(\hat{#1})}
\def\chbch#1#2{\langle#1\bar #2\rangle}

\newcommand{\beq}{\begin{equation}}
\newcommand{\eeq}{\end{equation}}

\title{
Sommerfeld effect for continuum gamma-ray spectra from Dark Matter annihilation}
\author{Barbara J\"ager\orcidlink{0000-0002-7005-9193}, Martin Vollmann\orcidlink{0000-0003-2193-8922}}

\affil{
Institute for Theoretical Physics, University of T\"ubingen,
Auf~der~Morgenstelle~14, 72076 T\"ubingen, Germany 
}

\begin{document}
\normalfont
\maketitle

\begin{abstract}
We present a calculation of the  continuum part of the gamma-ray spectra resulting from Dark Matter annihilation in the framework of the MSSM taking into account Sommerfeld effects. Concentrating on pure wino and pure higgsino scenarios we compare our calculation to existing work and explore the numerical impact of the features not captured by previous approximative descriptions. We find that, in particular for large neutralino masses, when the Sommerfeld enhancement is very large, chargino-antichargino annihilation processes, which have not been considered before, lead to sizable differences with respect to existing calculations.  
In scenarios with neutralinos in the intermediate-mass range, we find that the role of the charginos is crucial in the endpoint regime.
Our calculation provides the currently most accurate prediction for the continuum gamma-ray spectra. 
\end{abstract}

\section{Introduction}
Understanding the nature and origin of Dark Matter (DM) represents one of the great challenges of contemporary research in cosmology, astro- and particle physics. While there is little doubt in the community on the very existence of such a form of matter that does not interact electromagnetically with ordinary matter, the Standard Model of particle physics (SM) does not offer an obvious candidate particle to account for it. 

Promising DM candidates are provided by models of physics beyond the SM (BSM) featuring weakly interacting massive particles (WIMPs) with masses in the range of a few hundred~GeV to a few~TeV.  It is assumed that after a period of  equilibrium between thermal WIMP production and annihilation in the early Universe a freeze-out of their number density occurred when the expansion rate of the Universe became larger than the relevant interaction rate. The measured abundance of DM is compatible with the existence of a WIMP at the electroweak (EW) mass scale. 

A natural framework for WIMPs is provided by supersymmetric theories that complement the particle spectrum of the SM by so-called superpartners with spin differing by one half. The breaking of supersymmetry (SUSY) allows these particles to acquire masses different from their SM partners. In many SUSY models the lightest stable particle (LSP) represents a viable DM candidate. A prime example of such a model is constituted by the Minimal Supersymmetric Standard Model (MSSM).  The MSSM contains a minimum number of SUSY particles, i.e.\ one superpartner for each SM particle apart from the Higgs boson. The Higgs sector has to be extended, resulting in five physical bosonic Higgs states and the same number of fermionic  higgsinos.  Form a theoretical point of view, the MSSM is appealing not only because of its underlying symmetry structure, but also by ensuring the  unification of gauge couplings, naturalness, and providing a DM candidate. In most realizations of the MSSM the lightest neutralino, which effectively is a mixture of bino, wino and higgsino components, is considered stable and thus constitutes a viable WIMP DM candidate. 

Great hopes for the discovery of SUSY particles were pinned on the CERN Large Hadron Collider (LHC) with the highest collision energies ever achieved on Earth by a proton accelerator. Dedicated searches by the  ATLAS and CMS experiments put limits on the MSSM parameter space, but still leave room for the existence of SUSY particles, in particular in so-called   {\em split} scenarios where the LSP is much heavier than the scale of EW symmetry breaking \cite{Arkani-Hamed:2004ymt,Giudice:2004tc,Wells:2004di}. In such scenarios neutralino mixing effects are typically suppressed and the DM candidate is part of a ``pure''  EW multiplet \cite{Arkani-Hamed:2006wnf}. Such scenarios are appropriately referred to as {\em higgsino} or {\em wino} type.  Due to the  large neutralino masses these DM models predict they remain elusive 
to searches at colliders and direct detection experiments.

Depending on the underlying model, DM candidates are supposed to be produced in pairs or in association with accompanying particles in collisions of SM particles~\cite{ATLAS:2021moa,CMS:2020bfa}.  If the masses of the DM particles are large and/or their interactions with SM particles feeble, the corresponding production rates are low.  The same limitation applies to direct detection experiments that aim to identify the recoil of a DM particle off a nuclear target~\cite{XENON:2023sxq,LZ:2022ufs,CRESST:2019jnq,XENON:2017lvq}. 

Promising alternatives for DM searches in such scenarios are constituted by indirect detection strategies aiming at identifying annihilation signatures of DM particles \cite{Jungman:1995df}.  In particular, when such annihilation processes are accompanied by photon emission the resulting gamma ray spectra feature a very characteristic peak structure 
at the kinematical endpoint. The shape and intensity of such spectra can be heavily influenced by the so-called \emph{Sommerfeld effect}. This, in turn, can be exploited to overcome severe limitations of indirect searches due to large uncertainties in  the distribution of DM in the inner galaxy and omni-present astrophysical backgrounds. The Sommerfeld effect is ubiquitous in annihilation processes involving non-relativistic particles that can exhibit long-range interactions. This phenomenon applies to MSSM neutralinos, where the long-range interactions are mediated by electroweak bosons.

A plethora of experiments and astrophysical observations has been devised to make use of this detection strategy.
Particularly interesting for indirect neutralino searches are various observatories. 
These include the space telescope Fermi-LAT \cite{Fermi-LAT:2009ihh} which is suitable for neutralino searches in the mass range of $\ord(1)-\ord(100)$~GeV. Additional information is coming from the currently operating Imaging Air Cherenkov Telescopes H.E.S.S. \cite{HESS:2006fka}, VERITAS \cite{VERITAS:2006lyc}, MAGIC \cite{MAGIC:2014zas}, along with their next-generation counterparts CTA \cite{CTAConsortium:2010umy} and LHAASO \cite{LHAASO:2019qtb}. 
The water Cherenkov telescope HAWC \cite{DeYoung:2012mj} further enriches this list. 
These Cherenkov telescopes are particularly suited for neutralino searches with somewhat heavier masses in the range of $\ord(0.1)-\ord(100)$~TeV.

Specific SUSY searches using gamma-ray observations have been conducted in Refs.~\cite{Beneke:2016jpw,GAMBIT:2017zdo,GAMBIT:2017snp,Hryczuk:2019nql,Ellis:2022emx,Arcadi:2022hve,Foster:2022nva} while in Refs.~\cite{Fermi-LAT:2015att,MAGIC:2016xys,HAWC:2017mfa,HESS:2018kom,VERITAS:2017tif,Michailidis:2023pkd} similar searches and predictions  have been presented for more generic WIMP hypotheses.

From the theoretical point of view, exploiting the full potential of this search strategy requires a quantitative understanding of the Sommerfeld effect.  In the context of the \emph{full} MSSM, work to that effect has been limited. In particular, Refs.~\cite{Beneke:2016jpw,Hryczuk:2019nql} are, to our knowledge, the only papers including the Sommerfeld enhancement in their MSSM analyses. In these studies, however, the incorporation of the Sommerfeld effect is achieved through the  matching of exclusive 2-to-2 Sommerfeld-resummed neutralino-annihilation cross sections. This approach has some caveats and, as we will argue,  cannot account for several pivotal phenomenological aspects that can become crucial for future analyses.

We aim to fill that gap with this work. In particular, we compute \emph{all} neutralino and chargino annihilation cross sections into three-body final states that are relevant for obtaining the continuum gamma-ray spectra. We obtain these annihilation cross sections in analytical form for arbitrary spin and helicity combinations of the final-state particles.

While our results are valid for generic MSSM parameter sets, for our numerical discussion we focus on the pure wino and higgsino limits. We find that the impact of the Sommerfeld effect on the continuum spectrum is sizable even when the neutralino mass is of the order of a few hundred~GeV. Our results show that, besides the Sommerfeld effect being very large~\cite{Hisano:2003ec}, the chargino contribution dominates for the intermediate to the very high energy part of the gamma-ray spectrum.  This aspect has not been captured by previous computations.

The paper is structured as follows: In Sec.~\ref{sec:ann} we briefly review the basic theoretical aspects of indirect DM detection using gamma rays. We then move on to the discussion of the Sommerfeld effect in the context of SUSY and the methods we used for the calculation of annihilation cross sections  in Sec.~\ref{sec:somm}. In Sec.~\ref{sec:res} we discuss our numerical results in the context of pure wino and higgsino scenarios, and we then conclude in Sec.~\ref{sec:conclusions}.  Conventions and some technical aspects of our work are discussed in App.~\ref{app:conventions} and App.~\ref{app:phase-space}.  
\section{The gamma-ray emission spectrum resulting from neutralino annihilation}
\label{sec:ann}
DM halos of nearby galaxies feature characteristic gamma-ray emission signals with an associated  flux  
\begin{equation}
\frac{\der\Phi}{\der E_\gamma} = \frac1{8\pi\Mdm^2}J_{\rm obs}\left\langle\frac{\der(\sv)}{\der E_\gamma} \right\rangle\ , 
\end{equation}
where $J_{\rm obs}$ is the astrophysical ``$J$'' factor \cite{Bergstrom:1997fj} for a given observed region and $\langle\der(\sv)/\der E_\gamma\rangle$ is the velocity-averaged annihilation cross section of two DM candidate particles $\chi$ of mass $m_\chi$ into gamma rays. With some rare exceptions the $J$~factors are independent of the gamma-ray energy $E_\gamma$, and are thus irrelevant for the description of the spectral properties of the DM gamma-ray signals, which are the focus of this work. We refer the reader interested in specific $J$~factors to Refs.~\cite{Pieri:2009je,McMillan:2011wd,Iocco:2011jz} for the Milky Way and to Refs.~\cite{Martinez:2013els,Geringer-Sameth:2014yza,Gavazzi:2009kf} for other astrophysical environments.

Here, however, we concentrate on the DM annihilation cross section $\der(\sv)/\der E_\gamma$ with the DM candidate particle being constituted by the lightest neutralino $\chi_1^0$ of the MSSM. This annihilation cross section has a kinematic endpoint at $E_\gamma\simeq\Mdm$\footnote{We use natural units: $\hbar=c=1$.}. At this energy, the spectrum features a \emph{quasi-monochromatic gamma-ray} line with a (natural) broadening of $\mathcal O(v^2)$, where $v$ is the average speed of the neutralinos in the DM halos of interest ($v\approx 10^{-3}$ in the Milky Way). The monochromatic nature of the line is due to the two-body kinematics of the neutralino-annihilation process into photons,  $\chi_1^0\chi_1^0\to \gamma\gamma$. On top of the gamma-ray spectral line, the endpoint spectrum of neutralino annihilation involves a $Z$ resonance (from the $\chi_1^0\chi_1^0\to \gamma Z^*$ process) with a natural width of $\Gamma_Z/m_Z\sim 0.03$. Both the $\gamma\gamma$ and the $\gamma Z$ contributions at the endpoint of the spectrum are, in principle, loop suppressed. However, the narrow width of the $Z$ resonance and the significant influence of long-range interactions between the neutralinos and the charginos due to the Sommerfeld effect make these features highly intriguing in terms of detectability.

\begin{figure}[t!]
	\begin{center}
	\includegraphics[width=.6\linewidth]{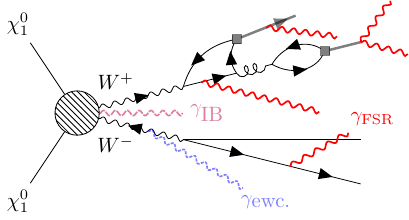}
	\end{center}
	\caption{\label{fig:continuum} Illustrative Feynman diagram for the continuum contribution to the gamma-ray emission spectrum produced by neutralino annihilation. Red lines depict final-state photon radiation, blue lines soft/collinear electroweak radiation, and purple lines internal bremsstrahlung. }
\end{figure}

The remaining part of the spectrum, \emph{the continuum}, is generated by the sum of all processes where neutralinos annihilate into a gamma-ray photon in association with a multiparticle configuration ``$X$'', i.~e.\ $\chi_1^0\chi_1^0\to \gamma+X$. At leading order (LO) in the electroweak coupling, $X$ consists of two SM particles. For clarity, we use a superscript in order to differentiate such a two-body state $X^{(2)}$ from the more complex sub-states $X$ that can occur in general, and that are illustrated by Fig.~\ref{fig:continuum}. 
%

In many simulations, such continuum configurations are approximated  by matching fixed-order computations for two-particle production processes to parton-shower programs such as {\tt Pythia}~\cite{Bierlich:2022pfr} or \texttt{Herwig}~\cite{Bellm:2015jjp}. In this approach gamma-ray spectra from WIMP annihilation are obtained using the parton-shower approximation formula
\begin{equation}
\label{eq:fsr}
\frac{\der(\sv)}{\der E_\gamma}\approx
\sum_{X^{(2)}}(\sv)_{X^{(2)}}
\frac{\der N_{X^{(2)}\to\gamma}^\textrm{MC}}{\der E_\gamma}\ ,
\end{equation}
where, given a specific WIMP model, the coefficients $(\sv)_{X^{(2)}}$ are the (tree-level) cross sections for the annihilation of two DM candidate particles into two SM particles (e.g.\ $ X^{(2)}=b\bar b, \tau^+\tau^-, W^+W^-, \ldots$), while the model-independent functions $\der N_{X^{(2)}\to\gamma}^\textrm{MC}/\der E_\gamma$  for the gamma-ray emission off the SM particles are obtained from the aforementioned Monte-Carlo event generators.\footnote{In the leading-logarithmic approximation of the aforementioned event generators, the fragmentation of the two SM particles is considered independently. This implies that $\der N_{X^{(2)}\to\gamma}^\textrm{MC}/\der E_\gamma$ in Eq.~\eqref{eq:fsr} actually denotes the  sum 
of two fragmentation functions -- one for each leg of primaries in Fig.~\ref{fig:continuum}. 
}  
These functions are available in specialized software packages such as \texttt{DarkSUSY}~\cite{Bringmann:2018lay}, PPPC~\cite{Cirelli:2010xx} or \texttt{MicrOMEGAs}~\cite{Belanger:2010gh}. 

As a diagrammatic visualization, in Fig.~\ref{fig:continuum} we show an example of a Feynman diagram for the continuum contribution to the gamma-ray emission spectrum produced by neutralino annihilation. From that particular diagram only those photon lines that are emitted from final-state legs are accounted for by the early implementations of Eq.~\eqref{eq:fsr} (e.~g.\ \texttt{DarkSUSY} versions before {\tt v5}  \cite{Gondolo:2004sc}). Such contributions are termed {\em final-state radiation} (FSR). 
Soft/collinear electroweak radiation effects have also been included in the literature (see, e.g.,~Ref.~\cite{Ciafaloni:2010ti}). 
In particular, the fragmentation functions provided by the PPPC code include these corrections. Newer versions of \texttt{DarkSUSY}  instead provide a more complete picture for the neutralino-annihilation photon spectrum at the expense of losing the model-independence of Eq.~\eqref{eq:fsr}. In addition to pure FSR calculations they capture potentially dominant processes such as the so-called {\em internal bremsstrahlung} (IB) \cite{Bergstrom:1989jr,Flores:1989ru,Baltz:2002we,Bringmann:2007nk,Barger:2009xe}, which is absent in the PPPC approach~\cite{Cirelli:2010xx}. This is achieved by computing the full fixed-order  $\neu1\neu1\to\gamma+X^{(2)}$ annihilation cross section for a given state $X^{(2)}$ ($X^{(2)}=W^+W^-$ in the sample diagram of Fig.~\ref{fig:continuum})  and matching it to a parton-shower program  while carefully subtracting redundant terms in order to avoid double counting, see e.~g.~Ref.~\cite{Bringmann:2007nk}.  In particular, IB becomes important in those cases where the otherwise helicity-suppressed annihilation of non-relativistic Majorana particles into a particle-antiparticle pair of light fermions becomes sizable once radiation effects are accounted for. It has been even observed that in some WIMP models, IB gives the dominant contribution to the gamma-ray spectrum in the medium-to-high energy regime (see Refs.~\cite{Bell:2011if,Garny:2011cj,Bringmann:2017sko} for more details).
\section{Sommerfeld effect}
\label{sec:somm}
\begin{figure}[t!]
	\begin{center}
		\includegraphics[width=.7\linewidth]{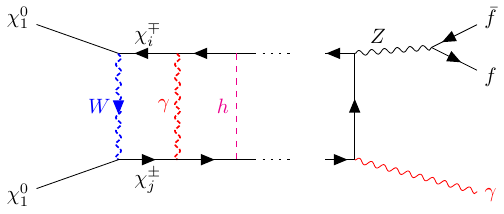}
	\end{center}
	\caption{\label{fig:SEmaindiag} Prototypical ladder-like Feynman diagram contributing to the Sommerfeld effect.
}
\end{figure}

The previous discussion evidences that great progress have been achieved in understanding the continuum part of the gamma-ray spectrum from neutralino annihilation. By combining fixed-order computations with Monte-Carlo event generators (see Eq.~\eqref{eq:fsr}), a relatively adequate theoretical picture of the WIMP gamma-ray spectrum  can be obtained. This picture, however, fails at capturing \emph{virtual} (loop) effects which, as we will see below, can become crucial. In particular, multi-loop corrections such as the one depicted in Fig.~\ref{fig:SEmaindiag} can have an enormous impact on the annihilation cross sections of heavy neutralinos~\cite{Hisano:2002fk,Hisano:2003ec,Hisano:2004ds}, resulting in enhancement factors of several orders of magnitude.
%
%
%
%

The primary reasons for this phenomenon are (A) the non-relativistic nature of the initial state consisting of two neutralinos, (B) $t$-channel interactions between them mediated by the gauge and Higgs bosons of the theory (see sketch in Fig.~\ref{fig:SEmaindiag}), and (C) the fact that these exchange bosons are lighter than the annihilating neutralinos. These features make a quantum-mechanical approach in terms of static potentials the most appropriate in order to account for the self-interactions of the neutralinos prior to their annihilation. In more formal terms, the computation of Feynman diagrams such as the one depicted in Fig.~\ref{fig:SEmaindiag} will inevitably yield terms that are (parametrically) 
\footnote{Note that the parameters $v\,\Mdm$, $m_Z$, $m_W$, $m_h$ are assumed to be of the same order of magnitude, and much smaller than $\Mdm$.}
of $\ord(\alpha^n\Mdm^{n}/m_W^{n})$, where $n$ is the number of loops in the given ladder-like diagram and $\alpha$ denotes the fine structure constant. Therefore, since we assume that $m_W\ll\Mdm$, the contribution of each of these diagrams is large and cannot be neglected. Rather all such terms need to be systematically \emph{resummed}.

The resummation of these diagrams can be performed consistently in the context of non-relativistic effective field theories (NREFTs)~\cite{Bodwin:1994jh} (in the context of MSSM neutralino annihilations,  see e.~g.~Refs.~\cite{Beneke:2012tg,Hellmann:2013jxa,Beneke:2014gja}) \emph{independently of the exclusiveness} of the final state. For the particular case of neutralino annihilation into photons, the resulting annihilation cross section is given by~\cite{Beneke:2018ssm,Beneke:2022pij}
%
\begin{equation}
\label{eq:SEg+X}
\frac{\dif{}(\sv)}{\dif \Eg}=2\sum_{IJ}S_{IJ}\left[\frac{\dif{} \tsv}{\dif\Eg}\right]_{IJ}\,, 
\end{equation}
%
where $S_{IJ}$ is the matrix of the so-called \emph{Sommerfeld factors},  and we will refer to $\dif{} \tsv/\dif\Eg$ as the \emph{annihilation matrix} for the Sommerfeld-corrected $\neu1\neu1\to\gamma+X$ process. The pair indices $I,J$ denote all possible neutral combinations of neutralino-neutralino and chargino-antichargino pairs within the MSSM \cite{Beneke:2014gja}. 
There are 14 such combinations: 
\begin{align}
\{(\neu1\neu1), (\neu1\neu2), (\neu1\neu3), (\neu1\neu4), (\neu2\neu2){}&{}, (\neu2\neu3), (\neu2\neu4), (\neu3\neu3),
(\neu3\neu4), (\neu4\neu4),\nonumber\\
(\cha1\acha1), (\cha1\acha2){}&{}, (\cha2\acha1), (\cha2\acha2)\}\,. 
\label{eq:schrbasis}
\end{align}
We compactly denote each pair index $K$ by nested particle indices distinguishing between neutralino and chargino states using either round or angular brackets, i.e.\ 
$K=\{\nn{ij}\}$, where $\nn{ij}=\nn{11},\, \nn{12},\ldots$ for the four neutralinos ($i,j=1,\ldots,4$) and 
$K= \{\chbch xy\}$, with $\chbch{x}{y}= \chbch11, \chbch12, \ldots$ for the two charginos ($x,y=1,2$). 
Note that while the neutralino states satisfy $\nn{ji}=\nn{ij}$, the chargino states are not exchange symmetric ($\chbch yx\neq\chbch xy$).

\begin{figure}[t!]
	\begin{center}
		\includegraphics[width=.8\linewidth]{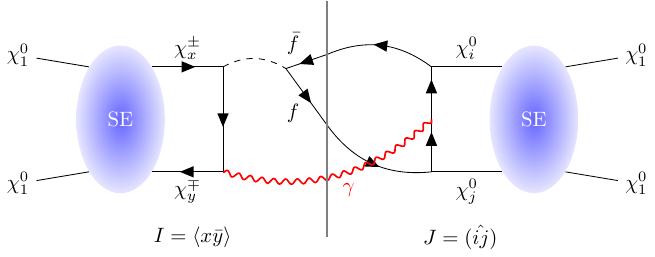}
	\end{center}
	\caption{\label{fig:sommfact} Sketch of a typical term in the Sommerfeld resummation formula of Eq.~\eqref{eq:SEg+X} for the  representative  $\neu1\neu1\to\gamma+ f\bar f$ process. 
}
\end{figure}

Figure~\ref{fig:sommfact} 
captures the essence of Eq.~\eqref{eq:SEg+X} in the representative  $\neu1\neu1\to\gamma+f\bar f$ process. 
In that particular example, the corresponding (interference) term of the equation with $I=\chbch xy$ and $J=\nn{ij}$ is shown. The diagram explicitly illustrates that both the Sommerfeld factors and the annihilation matrix elements can be expressed as products of quantities corresponding to the intermediate states $I$ and $J$, respectively. In the calculation of the annihilation matrix elements, the amplitudes of the  $\cha x\acha y\to\gamma+ f\bar f$ and the (complex conjugated) $\neu i\neu j\to\gamma+f\bar f$ processes enter. For the computation of the $S_{IJ}$ factors quantum mechanical wave functions evaluated at the spatial  origin have to be multiplied with their complex conjugates. These wavefunctions describe the non relativistic $\neu1\neu1\xrightarrow{\rm SE}\cha x\acha y$ and (complex conjugated) $\neu1\neu1\xrightarrow{\rm SE}\neu i\neu j$ transitions as we briefly discuss in the following.

%
%
\subsection{Computation of the Sommerfeld factors}
Up to corrections of $\ord(v^2)$ and $\ord(m_W^2/\Mdm^2)$, the resummation of ladder-like diagrams such as the ones shown in Fig.~\ref{fig:SEmaindiag}  requires solving the following matrix Schr\"odinger equation
\footnote{Here we are implicitly exploiting  that in the partial-wave decomposition of the non-relativistic wave functions the contribution of states with non-vanishing orbital momentum ($\ell\neq0$) is suppressed by factors of order $\ord(v^{2\ell})$. We thus only retain contributions with $\ell=0$. }
for the multicomponent function $u_{IK}(r)$ \cite{Beneke:2014gja} 
\begin{align}
\label{eq:schr}
-\frac1{\Mdm}\frac{\der^2u_{IK}}{\der r^2}(r)+\sum_J\left[\Delta M+V(r)\right]_{IJ}u_{JK}{}&{}(r) = \Mdm v^2u_{IK}(r)\, ,\\
\label{eq:schrinitial}
u_{IK}(r=0)=0\,,\quad\frac{\der u_{IK}(r=0)}{\der r}{}&{}=\delta_{IK}\,.
\end{align}
The Sommerfeld factors $S_{IJ}$ in Eq.~\eqref{eq:SEg+X} are then  given by 
\beq
S_{IJ}=U^{*\,-1}_{I\nn{11}}U^{-1}_{J\nn{11}}\,,
\eeq
where 
\beq
U_{IK}=\lim_{r\to\infty}\left[\e^{ik_Ir}\left(\frac{\der u_{IK}(r)}{\der r}-ik_Iu_{IK}(r)\right)\right]\,,
\eeq
$k_I\equiv\sqrt{\Mdm^2 v^2-\Mdm (M_I-2\Mdm)}$ and $M_I$ is the sum of the masses of the non-relativistic ``$I$'' state, e.~g. $M_{\chbch 12}=m_{\chapm1}+m_{\chapm2}$. 

The potential matrix in terms of the ``coupling matrices'' $\hv{B}_{IJ}$ is given by
\beq
\label{eq:mssmpot}
V_{IJ}(r)=
\begin{pmatrix} 
\displaystyle\sum_{B^0=\gamma,Z,h,H^0,A^0}\hv{B^0}_{\nn{ij},\nn{kl}}\frac{\e^{-m_{B^0}r}}r & \displaystyle\sum_{B^+=W,H^+}\hv{B^+}_{\nn{ij},\chbch zw}\frac{\e^{-m_{B^+}r}}r \\ 
\displaystyle\sum_{B^+=W,H^+}\hv{B^+}_{\chbch xy,\nn{kl}} \frac{\e^{-m_{B^+}r}}r  & \displaystyle\sum_{B^0=\gamma,Z,h,H^0,A^0}\hv{B^0}_{\chbch xy,\chbch zw}\frac{\e^{-m_{B^0}r}}r 
\end{pmatrix}\,,
\eeq
while the  ``mass-splitting'' matrix $\Delta M$ in the basis of Eq.~\eqref{eq:schrbasis} reads
$$\Delta M=\diag(0,m_{\neu1}+m_{\neu2}-2\Mdm,\ldots,m_{\chapm1}+m_{\chapm2}-2\Mdm,2\,m_{\chapm2}-2\Mdm)\,.$$ 

The Coulomb part of the potential ($B=B^0=\gamma$) satisfies 
\beq
\hv{\gamma}_{\nn{ij},\nn{kl}}=0\,, \quad \textrm{and} \quad
\hv{\gamma}_{\chbch xy,\chbch zw}=\alpha\,\delta_{xz}\delta_{yw},
\eeq 
where $\alpha$ is the fine-structure constant. The remaining coupling matrices depend on the MSSM parameters through the neutralino and chargino  mixing matrices defined in Eqs.~\eqref{eq:mssmneumatrix}-- \eqref{eq:mssmchamatrix} of App.~\ref{app:conventions}. These had been already obtained in Ref.~\cite{Beneke:2014gja}, and we re-derived them using the conventions of Ref.~\cite{Hahn:2001rv} that we use throughout this work. In the Feynman gauge the entries of these coupling matrices read 
\begin{eqnarray}
\hv{Z}_{\nn{ij},\nn{kl}} &=& \frac1{\sqrt2^{\delta_{ij}}}\frac1{\sqrt2^{\delta_{kl}}}\frac{\alpha}{s_W^2}\left[v_{ik}^{Z\,(0)}v_{jl}^{Z\,(0)\,*} - 3\,a_{ik}^{Z\,(0)}a_{jl}^{Z\,(0)\,*}+s_{ik}^{G_Z\,(0)}s_{jl}^{G_Z\,(0)\,*} + (k\leftrightarrow l)\right], 
\label{eq:nnZnnpot}
\\
\hv{Z}_{\chbch xy, \chbch zw} &=& \frac{\alpha}{s_W^2}(v^Z_{xz}v^{Z\,*}_{yw} - 3\,a^Z_{xz}a^{Z\,*}_{yw}+s_{xz}^{G_Z}s_{yw}^{G_Z\,*}),
\label{eq:ccZccpot}\\
\hv{W}_{\nn{ij},\chbch zw} &=& \frac1{\sqrt2^{\delta_{ij}}}\frac{\alpha}{s_W^2}\left[v_{iz}^{W}v_{jw}^{W\,*} - 3\,a_{iz}^{W}a_{jw}^{W\,*}+s_{iz}^{G_W}s_{jw}^{G_W\,*} + (z\leftrightarrow w)\right],
\label{eq:nnWccpot}
\end{eqnarray}
for the vector mediators $Z$ and $W^\pm$ and 
\begin{eqnarray}
\hv{S}_{\nn{ij},\nn{kl}} &=& \frac1{\sqrt2^{\delta_{ij}}}\frac1{\sqrt2^{\delta_{kl}}}\frac{\alpha}{s_W^2}\left[s_{ik}^{S\,(0)}s_{jl}^{S\,(0)\,*} + (k\leftrightarrow l)\right], \\
\hv{S}_{\chbch xy, \chbch zw} &=& \frac{\alpha}{s_W^2}s_{xz}^Ss_{yw}^{S\,*}\,, \\
\hv{H^\pm}_{\nn{ij},\chbch zw} &=& \frac1{\sqrt2^{\delta_{ij}}}\frac{\alpha}{s_W^2}	\left[s_{iz}^{H^\pm}s_{jw}^{H^\pm\,*} + (z\leftrightarrow w)\right],
\label{eq:nnHccpot}
\end{eqnarray}
for the scalar and pseudoscalar  mediators, where $S$ collectively labels all physical neutral Higgs bosons ($S=h,H,A^0$) 
in the MSSM, and the would-be Goldstone bosons $G_{Z}$, $G_{W}$  associated with the $Z$ and $W$ bosons. 

Explicit expressions for all the coefficients $v^B_{IJ}$, $a^B_{IJ}$ and  $s^B_{IJ}$ can be found in App.~\ref{app:conventions}. We obtained these by using the {\tt Mathematica}~\cite{Mathematica} packages \href{http://www.feynarts.de/}{\tt FeynArts}~\cite{Hahn_2001} and {\tt FormCalc}~\cite{Hahn_1999,Hahn:2016ebn} with the MSSM model file~\cite{Hahn:2001rv}.  At the relevant perturbative order, our  potential agrees with the corresponding results of Ref.~\cite{Beneke:2014gja}.

\subsection{Computation of the annihilation matrices}
\label{sec:annmat}
In the previous section we briefly reviewed the NREFT methods that are necessary in order to properly incorporate the Sommerfeld effect in the continuum spectrum prediction. In particular, Eq.~\eqref{eq:SEg+X} provides us with an elegant prescription on how to deal with this problem. Given the Sommerfeld coefficients $S_{IJ}$ which can be computed by solving a system of Schr\"odinger equations, we need to determine the $[\der\tsv/\der\Eg]_{IJ}$ functions for every possible combination~$I,J$. The $IJ$ element of the annihilation matrix is defined by
\begin{equation}
\label{eq:annmat}
\left[\frac{\dif{}\tsv}{\dif \Eg}\right]_{IJ}=\frac1{(\sqrt2)^{\id(I)+\id(J)}}\frac1{4\Mdm^2}\sum_X\int\dif\Pi'_{\gamma+X}\delta\left(\Eg-\Eg'\right)\amp^{(l,s)=(0,0)}_{I\to\gamma+X}\amp^{(l,s)=(0,0)\,*}_{J\to\gamma+X}\ ,
\end{equation}
where the identical-particle index $\id(K)=\delta_{ij}$ for neutralino pairs [$K=\nn{ij}$] or 0 for chargino-antichargino pairs [$K=\chbch xy$].  $\dif\Pi'_{\gamma+X}$ is the phase-space integration element for the $\gamma+X$ final state, and $\amp^{(l,s)=(0,0)}_{K\to\gamma+X}$ is the amplitude for the \emph{$s$-wave annihilation}
\footnote{A crucial point in Eq.~\eqref{eq:annmat} is the fact that both the $I$ and $J$ states  individually exhibit the same quantum numbers as the (physical) initial state of the system,  $\neu 1\neu 1$. In the partial-wave basis this is the $s$-wave state with vanishing orbital ($\ell=0$) and spin ($s=0$) quantum numbers  in virtue of the Majorana nature of the MSSM neutralinos, provided that $v\to0$.
} 
of the two-particle state $K$ into a $\gamma+X$ state. It is implicitly understood that the sum over $X$ includes all possible spin and helicity combinations of the particles that are produced in association with the gamma ray. 

In this article we compute the annihilation matrices of  Eq.~\eqref{eq:annmat} for  the production of a photon in association with any possible combination of two SM particles  $X=X^{(2)}$. Concretely, in the MSSM the only possible combinations are
\footnote{The processes $\neu1\neu1\to\gamma+ ZZ$, $\gamma +\gamma Z$ and $\gamma+\gamma\gamma$ are forbidden because of the Landau-Yang theorem. 
}
\begin{equation}
\label{eq:X}
X^{(2)}=\{W^+W^-,W^\pm H^\mp,H^+H^-,ZS,SS',f\bar f\}\ ,
\end{equation}
where $S$ and $S'$ are shorthand for two different neutral Higgs scalars in the MSSM  and $f\bar f$ denotes any fermion anti-fermion pair within the SM. There are $14\times(14+1)/2=105$ independent symmetric combinations of the initial-state indices (see Eq.~\eqref{eq:schrbasis}). Thus, for each one of the aforementioned combinations of $X^{(2)}$ 105 independent matrix elements have to be computed. Out of these matrix elements, in the literature only one is available \cite{Bringmann:2007nk},  corresponding to the neutralino-neutralino annihilation cross section with $I=J=\nn{11}$. In this work  we compute the annihilation matrices for the remaining 104 combinations of $I,J$. 

State-of-the-art software packages such as \href{http://www.feynarts.de/}{\tt FeynArts}~\cite{Hahn_2001} or {\tt FormCalc}~\cite{Hahn_1999,Hahn:2016ebn} are capable of computing  differential cross sections in analytical form for annihilation processes within complicated models such as the MSSM. However, the problem at hand requires some extra processing for the $s$-wave projection of the $I,J$ states and for computing the interference of amplitudes with different initial states.

We addressed these issues by obtaining raw amplitudes for the 2-to-3 scattering processes $\neu i\neu j\to\gamma+X^{(2)}$ of all possible $X^{(2)}$ states of Eq.~\eqref{eq:X} with arbitrary spin and helicity combinations using {\tt FeynArts 3.11}/{\tt FormCalc 9.8} and the built-in MSSM model file of Ref.~\cite{Hahn:2001rv}. We then processed these amplitudes in the framework of {\tt Mathematica} \cite{Mathematica} to obtain the desired interference contributions. 

\subsubsection{Annihilation matrices in the FSR approximation}
The calculation of the annihilation matrix  $\dif{}\tsv{}/\dif\Eg$ presented above is \emph{exact} at LO in the electroweak couplings. Thus, when multiplied with the Sommerfeld factors and matched with parton-shower simulations, the resulting prediction offers the most accurate description of the continuum gamma-ray spectrum from annihilating neutralinos up to the present day. However, an \emph{approximate} treatment that exploits the fact that the computation of the Sommerfeld factors is independent of the exclusiveness of the final state in the annihilation process, is also possible and has already been considered in Refs.~\cite{Beneke:2016jpw,Hryczuk:2019nql}. In that approach, Eq.~\eqref{eq:fsr} generalizes to 
\begin{equation}
\label{eq:fsr+se}
\frac{\der(\sv)}{\der E_\gamma}\approx
\sum_{X^{(2)}}(\sv)_{X^{(2)}}^\textrm{SE}
\frac{\der N^\textrm{MC}_{X^{(2)}\to\gamma}}{\der E_\gamma}\ ,
\end{equation} 
where, in analogy to Eq.~\eqref{eq:SEg+X}, the Sommerfeld-corrected 2-to-2 annihilation cross section is given by 
\beq
(\sv)_{X^{(2)}}^\textrm{SE}=2\sum_{IJ}S_{IJ}\,\tsv^{X^{(2)}}_{IJ}\,,
\eeq 
and, as just argued, the fragmentation functions remain the same as in Eq.~\eqref{eq:fsr}. Eq.~\eqref{eq:fsr+se} is sensitive to the different polarizations of the 2-body final states. To emphasize this feature, the polarization states can be indicated explicitly as 
$$X^{(2)}=\{W^+_TW^-_T,W^+_\odot W^-_\odot, W^\pm_LH^\mp,f_L\bar f_L,f_R\bar f_R,f_L\bar f_R,f_R\bar f_L,\gamma_T\gamma_T,\gamma_T Z_T,Z_TZ_T,Z_\odot Z_\odot,Z_\odot S,SS'\}\ ,$$
where the subscripts ``$T$'' and ``$\odot$'' refer to their transverse and longitudinal polarization components of the gauge bosons and the ``$L$'' and ``$R$'' subscripts denote the left- and right-handed chirality of the fermions, respectively. 
While the unpolarized cross sections for these 2-body states are already known \cite{Beneke:2012tg}, we obtain the polarized ones for the first time in this work. 
\section{Numerical results}
\label{sec:res}
In light of the inherent complexity of the MSSM, a thorough exploration of our calculations requires a dedicated study.
In this section, we focus on two limiting scenarios of the MSSM: the pure wino and the pure higgsino case. These models have been investigated intensively in the last several years (see e.~g.~\cite{Beneke:2016jpw,Hisano:2002fk,Cirelli:2005uq,Cohen:2013ama,Fan:2013faa,Rinchiuso:2020skh,Dessert:2022evk}).  We note that  even in the most generic MSSM parameter sets, the Sommerfeld effect has a significant impact on the associated indirect detection signals. 

The most general form of the softly broken MSSM introduces 105 parameters in addition to those present in the SM~\cite{Dimopoulos:1995ju}. These are interdependent given a specific supersymmetry breaking scenario.
Current experimental efforts to search for SUSY, however, concentrate on constrained versions of the MSSM such as the 18-parameter phenomenological MSSM (pMSSM) \cite{MSSMWorkingGroup:1998fiq} or the 4-parameter constrained MSSM (CMSSM) (see, e.g., Refs.~\cite{Kane:1993td,Ellis:2022emx}). The pMSSM, for example, treats the mass terms for gauginos ($M_1$, $M_2$, $M_3$), higgsino ($\pm|\mu|$), sfermions and trilinear couplings, as well as the ratio of the vacuum-expectation values of the two Higgs doublets ($\tan\beta$),  as independent parameters. 

In the pure wino and higgsino limits,  the sfermion masses as well as the trilinear couplings, the gluino mass parameter $M_3$, and $\tan\beta$ are assumed to be infinitely large. In the pure wino limit, in addition to integrating out the sfermions it is assumed that $M_1\to\infty$ and $|\mu|\to\infty$, leaving $M_2$ as the only relevant MSSM parameter. In the spirit of the minimal DM models discussed in Ref.~\cite{Cirelli:2005uq} (see also Refs.~\cite{Bottaro:2021snn,Bottaro:2022one} for a more recent study on minimal DM) the wino can be visualized as an SU(2) Majorana triplet which, after electroweak symmetry breaking, gives rise to one chargino and one neutralino degree of freedom. Similarly, the pure higgsino limit can be constructed by introducing an SU(2) Dirac doublet that results in two neutralinos and one chargino.
Note that the pure wino and higgsino scenarios do not provide satisfactory solutions to the hierarchy problem and the unification of gauge couplings, which were among the primary motivations for introducing SUSY. Moreover, for sub-TeV neutralino masses within the standard freeze-out hypothesis they predict  more DM than observed~\cite{Arkani-Hamed:2006wnf}. Nonetheless, these limiting scenarios serve as highly valuable toy models. For instance,  the dimension of the Sommerfeld and the annihilation matrices in Eq.~\eqref{eq:SEg+X} is reduced to two for the pure wino and to three for the pure higgsino model, respectively, thus significantly reducing the complexity of calculations and analyses as compared to the full MSSM. 

For the numerical results shown below as electroweak input parameters we choose the mass of the $Z$ boson,$m_Z=91.1876$~GeV, the Fermi constant, $G_\mu=1.1663788\times10^{-5}$~GeV${}^{-2}$, and the fine structure constant, $\alpha_0=1/137.035999180$ as quoted by the Particle Data Group~\cite{Workman:2022ynf}. From these input parameters,  the mass of the $W$~boson, $m_W=80.360$~GeV, and the Weinberg angle $\sin^2\theta_W=0.22338$ are obtained using  the electroweak relations  of Ref.~\cite{Workman:2022ynf}. In our numerical implementations we use the running QED coupling evaluated at the scale $m_Z$, $\alpha=1/128.93$. The masses of the $u$, $d$, $s$ quarks, the electron and muon, and of all  neutrinos  are neglected. For the $c$, $b$, $t$ quarks, the tau lepton, and the Higgs boson  we use the masses quoted in  Ref.~\cite{Workman:2022ynf}: $m_c=1.27$~GeV, $m_b=4.18$~GeV, $m_t=172.69$~GeV, $m_\tau=1.77686$~GeV, and $m_h=125.25$~GeV. Finally, for the relative speed of the annihilating particles we use the nominal value of $v=10^{-3}$.

For both, wino-like and higgsino-like models,  we use the one-loop expressions given in Refs.~\cite{Cirelli:2005uq,Ibe:2012sx} for the mass splitting between the chargino and the lightest neutralino, $m_{\chi_1^\pm}-m_{\chi_1^0}$. For the mass splitting between the two neutralinos  in the pure higgsino scenario we use $m_{\neu2}-m_{\neu1}=20$~MeV in all our examples, as done in Ref.~\cite{Beneke:2019gtg}. 

The numerical results that are reported below involve comparisons between different prescriptions and approximations. For simplicity we introduce the following acronyms: 
\begin{itemize}
\item \texttt{`fixed\_noSE'}:
Neutralino-pair annihilation into a $\gamma+X^{(2)}$ final state at tree-level with no Sommerfeld resummation at fixed order ($\ord(\alpha^3)$), as provided in Ref.~\cite{Bergstrom:2005ss}. Generally, $X^{(2)}$ denotes a two-particle state. In the pure wino and higgsino models, only $X^{(2)}=W^+W^-$ occurs. 
%
\begin{center}
	\includegraphics[width=.8\linewidth]{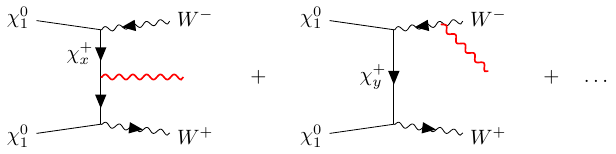}
\end{center}
%
\item \texttt{`FSR\_SE'}: 
Sommerfeld-corrected neutralino-pair annihilation into a $\gamma+X^{(2)}$ final state where only photon emission due to FSR in the soft/collinear approximation is taken into account.  
%
\begin{center}
	\includegraphics[width=.9\linewidth]{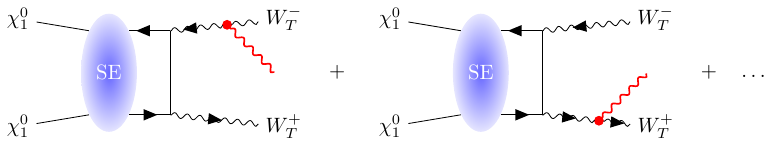}
\end{center}
%
For the pure wino and higgsino scenarios, the only possible two-body final state is $X^{(2)}=W_T^+W_T^-$. The spectrum that is obtained using this approximation is computed by using Eq.~\eqref{eq:fsr+se} with the LO fragmentation function of Ref.~\cite{Ciafaloni:2010ti} which in this case assumes the particularly simple form 
\begin{align}
\frac{\der N_{W_T^+W_T^-}}{\der\Eg}=\frac4{\Mdm}\,\frac{\alpha}{\pi}{}&{}\left\{
\frac{\Eg}{\Mdm-\Eg}\left[
\log\left(\frac{\Mdm-\Eg}{m_W}+\sqrt{\frac{(\Mdm-\Eg)^2}{m_W^2}-1}\right)\right.\right.\nonumber\\
+{}&{}\left.\left.\left(\frac{\Mdm-\Eg}{\Eg}+\frac{\Eg(\Mdm-\Eg)}{\Mdm^2}\right)\log\frac{2\,\Mdm}{m_W}
\right]
\right\}\,. 
\label{eq:fsrlo}
\end{align}

\item \texttt{`full\_SE'}: Neutralino-pair annihilation into a $\gamma+X^{(2)}$ final state including  continuum contributions and Sommerfeld resummation effects. In contrast to the two previous approaches, this computation receives contributions from all the two-particle combinations of Eq~.\eqref{eq:X}.
%
\begin{center}
	\includegraphics[width=.9\linewidth]{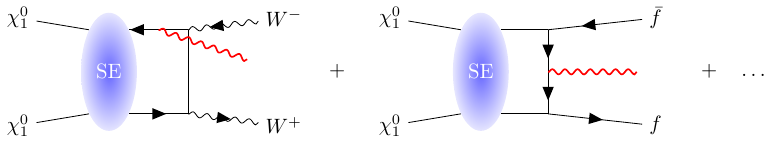}
\end{center}
%
\end{itemize}

%
\subsection{Scenarios with \boldmath$\Mdm=150$~GeV}
\begin{figure}
	\begin{center}
	\includegraphics[width=.7\linewidth]{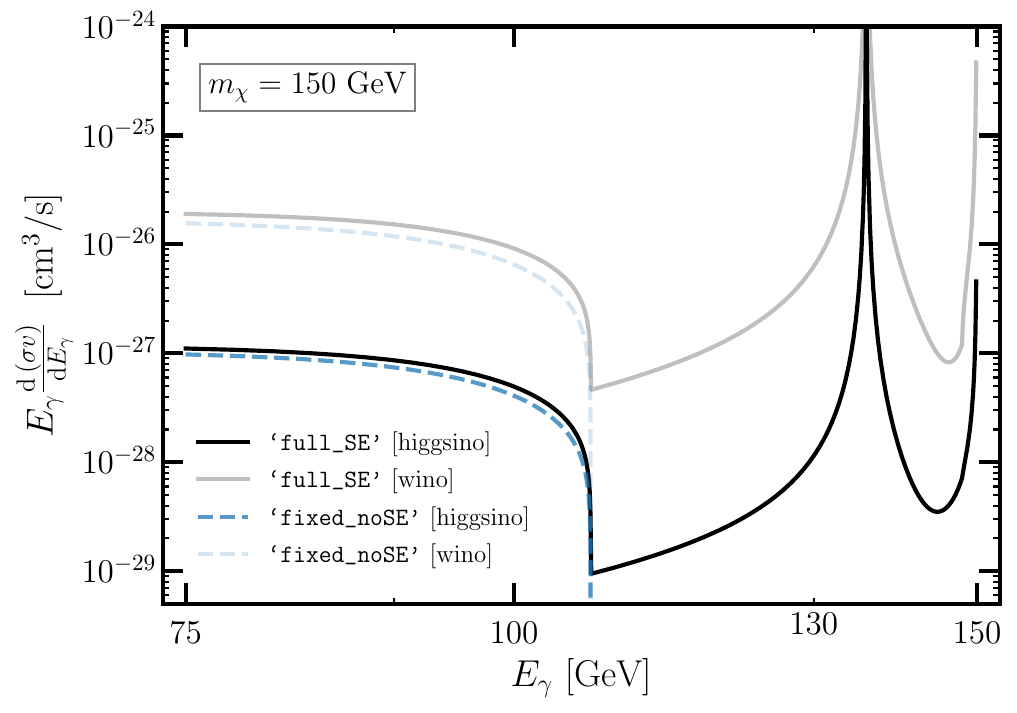} 
	\caption{\label{fig:mainplot} 
Gamma-ray spectrum of neutralino annihilation in the higgsino-like (black) and the wino-like (gray) scenarios with $\Mdm=150$~GeV within our full calculation (solid lines) and  the unresummed LO  computation of Ref.~\cite{Bergstrom:2005ss} (dashed lines). 
}
	\end{center}
\end{figure}

In Fig.~\ref{fig:mainplot}
%
%
%
%
we show the gamma-ray spectra of neutralino annihilation in these two scenarios for our full calculation including the Sommerfeld effect and compare it to the unresummed LO calculation of Ref.~\cite{Bergstrom:2005ss}. Note that the results of the \fsr{} approach are zero in the plot range due to the kinematic threshold in Eq.~\eqref{eq:fsrlo} at $\Eg\approx\Mdm-m_W\approx 70$~GeV. The comparison between the two remaining calculations illustrates impressively the relevance of the effects that have been neglected in the past, even when there are no large hierarchies between the masses of the lightest supersymmetric particles and the gauge bosons of the electroweak theory.
The most striking peculiarity of Fig.~\ref{fig:mainplot} is the appearance of a  resonance associated with a $Z$~boson contribution which is omitted by the ``na\"{\i}ve'' fixed order computation.  In particular, the fixed-order calculation has a clear cutoff from the kinematic threshold of the $\chi^0_1\chi^0_1\to W^+W^-\gamma$ process at the energy $E_\gamma=\Mdm-m_W^2/\Mdm$ (corresponding to about $107$~GeV in the scenarios of Fig.~\ref{fig:mainplot}).
The Sommerfeld-corrected continuum spectra \full{} that we compute here account for this $Z$-boson contribution by the inclusion of Feynman diagrams like  the one shown in Fig.~\ref{fig:SEmaindiag}. In such diagrams  fermion pairs can be created resonantly from a $Z$-boson in association with a photon with an energy that is larger than the aforementioned cutoff. The resulting peak of the $Z$ resonance is then at $E_\gamma=\Mdm-m_Z^2/(4\Mdm)$, amounting to about $136$~GeV in the scenarios of Fig.~\ref{fig:mainplot}.   

In addition to the $Z$ resonance, the spectral region we uncovered in this work exhibits further interesting features. The almost imperceptible kinks around the $E_\gamma =\Mdm$ endpoint (150~GeV in the scenarios of this section) are due to the kinematic thresholds in the $\chi^0_1\chi^0_1\to  f\bar f\gamma$ processes, where $ f$  denotes any charged fermion of the SM with non-vanishing mass. For instance, the  $\chi^0_1\chi^0_1\to b\bar b\gamma$ process is only possible, if $\Eg<\Mdm-m_b^2/\Mdm$ ($=149.88$~GeV in the current scenario).

We note that, as expected, the impact of the Sommerfeld effect in the spectral region covered by the original fixed-order computation of Ref.~\cite{Bergstrom:2005ss}, i.e.\ below the $W^+W^-$ threshold, is marginal. As the $\Mdm/m_W$ hierarchies become larger, the numerical impact becomes more pronounced, as we shall discuss below. Independent of the neutralino mass, however, annihilation cross sections and Sommerfeld factors from wino-like neutralinos are generically larger than in the higgsino-like case,  because the electroweak charges in the pure wino limit are a factor of two larger than those for the pure higgsino limit. For instance, the Born-level annihilation cross section of wino-like neutralinos into $W^+W^-$ pairs is sixteen times larger than the corresponding cross section for higgsino-like neutralinos.

%
\subsection{Scenarios with heavier neutralinos}
\begin{figure}
	\begin{center}
	\includegraphics[width=.8\linewidth]{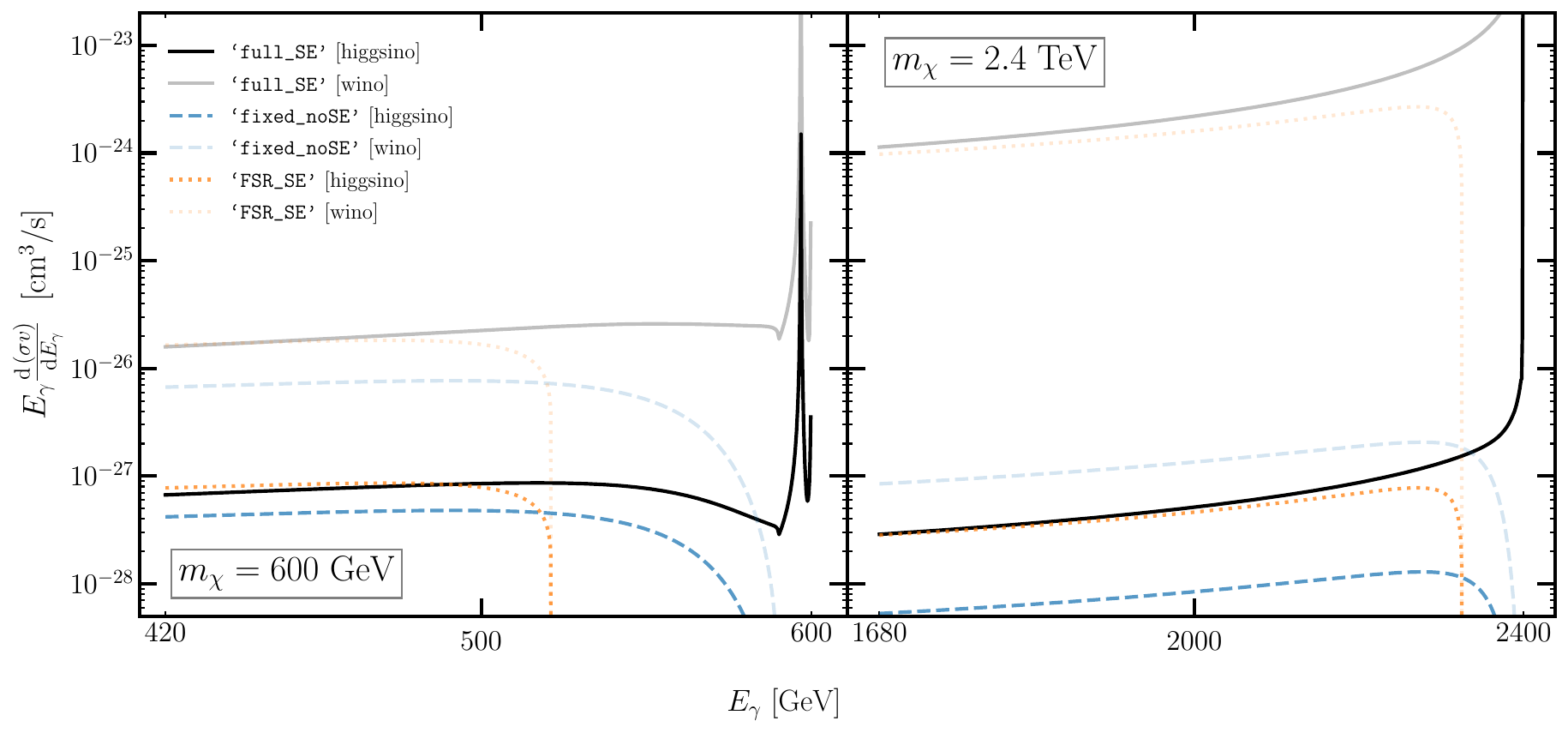}
	\includegraphics[width=.82\linewidth]{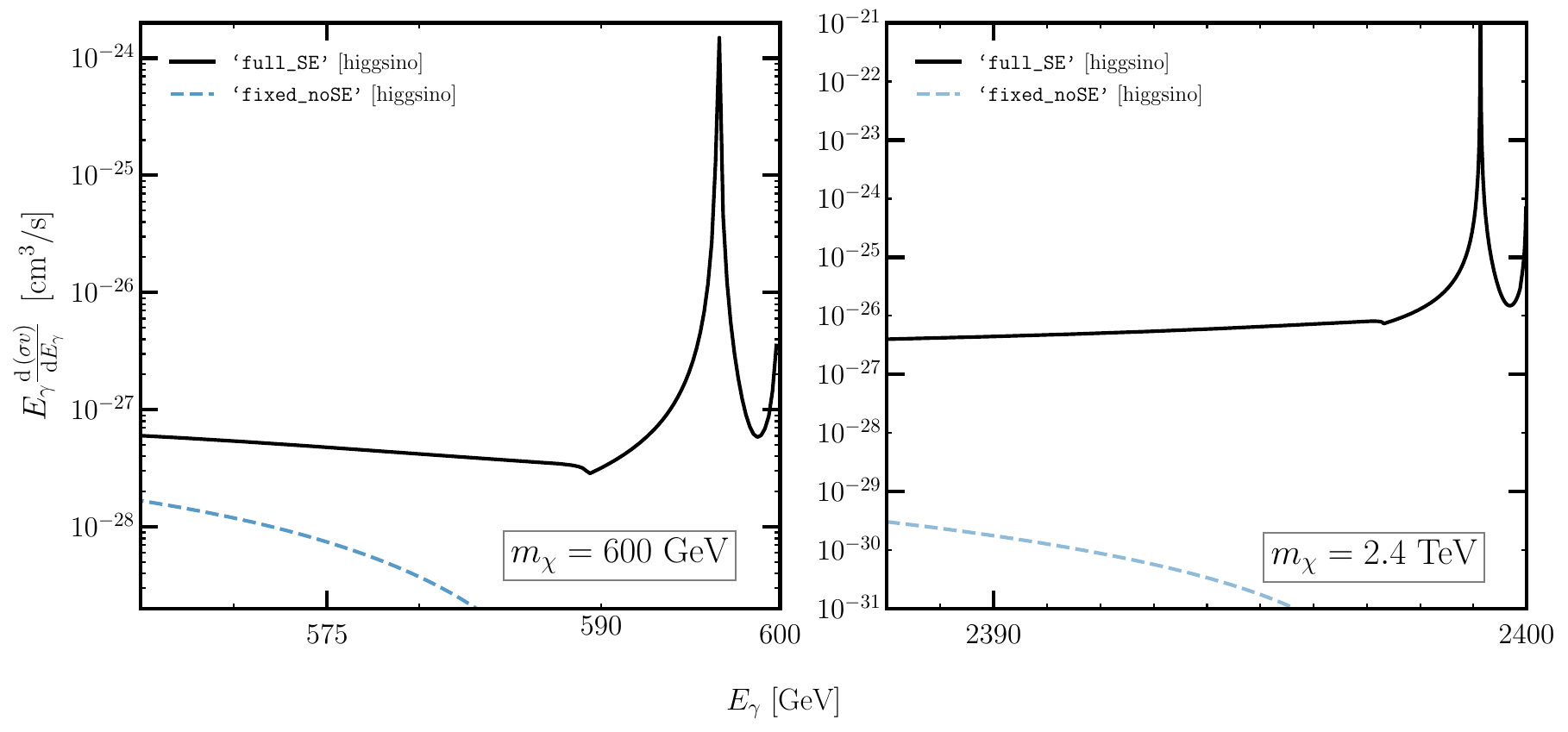}
	\includegraphics[width=.82\linewidth]{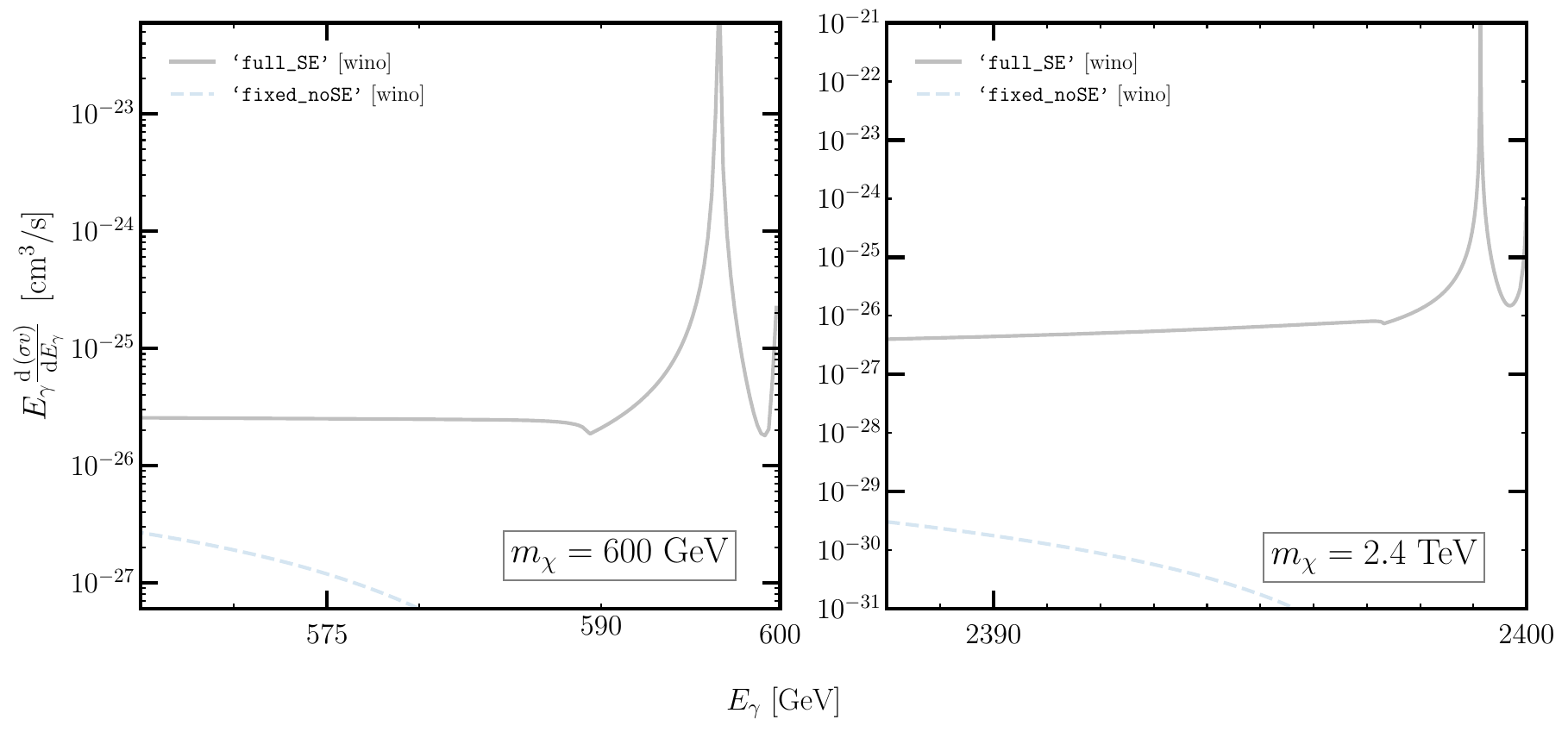} 
	\caption{\label{fig:doubleplot} 
	Upper panels: Gamma-ray spectrum of neutralino annihilation in the higgsino-like (black) and the wino-like (gray) scenarios with $\Mdm=600$~GeV (left) and  $\Mdm=2.4$~TeV (right) within our full calculation (solid lines),  the unresummed LO  computation of Ref.~\cite{Bergstrom:2005ss} (dashed lines), and the collinear-approximated computation of Ref.~\cite{Ciafaloni:2010ti} (dotted orange lines). 
	Middle panels:  Full calculation  (solid lines) and unresummed LO computation  (dashed lines) in the higgsino-like scenario close to the endpoint region. 
	Lower panels:  Full calculation  (solid lines) and unresummed LO computation  (dashed lines) in the wino-like scenario close to the endpoint region. }
	\end{center}
\end{figure}

Accounting for the Sommerfeld enhancement becomes more and more important as the mass of the neutralinos is increased. 
In Fig.~\ref{fig:doubleplot} we demonstrate this numerically for higgsino- and wino-like scenarios with a lightest neutralino of mass 600~GeV and 2.4~TeV, respectively. 
%
%
%
%

In each of these heavy neutralino scenarios, the predictions obtained within our full calculation show strong enhancements compared to the fixed-order computations. In the soft-photon part of the spectrum shown in Fig.~\ref{fig:doubleplot} we observe how well the FSR predictions \fsr{} match our improved predictions \full{} and how this agreement improves, as expected, when $\Mdm$ is increased. In particular, due to the model-independence of the splitting function of Eq.~\eqref{eq:fsrlo}, the ratios of the Sommerfeld-resummed calculations \fsr{} to the fixed-order results \fix{}
are roughly independent of $\Eg$ as long as  $\Eg\ll\Mdm$.
In the higgsino-like scenario these ratios amount to an enhancement by a factor of about $1.5$  for $\Mdm= 600$~GeV and of about 4.6 for $\Mdm=2.4$~TeV. In the wino-like scenario we find enhancement factors of $\sim1.9$ for $\Mdm= 600$~GeV and of $990$ for $\Mdm=2.4$~TeV.

This huge enhancement of the gamma-ray spectrum in the large-$\Mdm$ pure wino scenario is due to the resonant nature of the Sommerfeld effect (see e.~g.\ Fig.~1 of Ref.~\cite{Hisano:2003ec}). For particular neutralino masses the binding energy of a chargino-antichargino bound state is exactly zero, which ultimately results in a strong enhancement of the DM annihilation cross section. 
When varying the neutralino mass within our pure-wino scenario the enhancement factor reaches a first resonance at $\Mdm=2.29$~TeV.\footnote{The numerical value of the neutralino mass for which such resonance effects occur is very sensitive to the chargino-neutralino mass splitting, and also depends on NLO corrections to the non-relativistic potential, see e.~g.\  Ref.~\cite{Urban:2021cdu} and references therein. }  
It then decreases and increases again, until the second resonance  is reached at $\Mdm=  8.83$~TeV. Note that, as discussed e.~g.~in Ref.~\cite{Chun:2012yt}, depending on the SUSY parameters, the Sommerfeld resummation can lead to suppressed rather than enhanced annihilation rates.

As $\Eg$ increases, the situation becomes even more intriguing, requiring us to consider the relative impact of the various entries in  Eq.~\eqref{eq:SEg+X} for unraveling the underlying dynamics. More specifically, in the pure wino limit Eq.~\eqref{eq:SEg+X} can be decomposed into three terms:
%
%
\begin{eqnarray}
\frac{\dif{}(\sv)}{\dif \Eg}&=& 2\,S_{\nn{11}\nn{11}}\,\left[\frac{\dif{} \tsv}{\dif\Eg}\right]_{\nn{11}\nn{11}}+4\re\left\{S_{\nn{11}\chbch 11}\,\left[\frac{\dif{} \tsv}{\dif\Eg}\right]_{\nn{11}\chbch 11}\right\} \nonumber\\
{}&{}&{}2\,S_{\chbch 11\chbch 11}\,\left[\frac{\dif{} \tsv}{\dif\Eg}\right]_{\chbch 11\chbch 11}\ ,
\label{eq:winofact}
\end{eqnarray}
%
%
where a numerical evaluation of the Sommerfeld factors yields the results listed in Tab.~\ref{tab:snumbers}. 
%
%
\begin{table}[t!]
\begin{center}
\begin{tabular}{|c||c|c|c|}
\hline
{} & $\Mdm=150$~GeV & $\Mdm=600$~GeV & $\Mdm=2.4$~TeV \\
\hline
\hline
$S_{\nn{11}\nn{11}}$ & $1.011$ & $1.201$ & $251.4$ \\
\hline
$S_{\nn{11}\chbch 11}$ & $0.091$ & $0.449$ & $348.0$ \\
\hline
$S_{\chbch 11\chbch 11}$  & $0.008$ & $0.168$ & $481.8$\\
\hline
\end{tabular} 
\end{center}
\caption{
\label{tab:snumbers}
Elements of the Sommerfeld matrix in the pure-wino limit of the MSSM for the three neutralino masses considered in this section. }
\end{table}
%
In order to interpret the entries of the table, it is important to bear in mind that in the absence of the Sommerfeld effect $S_{IJ}=\delta_{I\nn{11}}\delta_{J\nn{11}}$. In the ``low mass'' scenario ($\Mdm=150$~GeV), this condition is  satisfied at the level of about 10~\%. This, for instance, explains why in Fig.~\ref{fig:mainplot}, the \fix{} and the \full{} computations are similar for $\Eg<107$~GeV. However, as noted before, the \fix{} calculation misses the $Z$ resonance effect  and all additional $f\bar f \gamma$ contributions resulting from the last term of Eq.~\eqref{eq:winofact}. As the neutralino mass increases, the numerical significance of this term as well as of the interference term (second term in Eq.~\eqref{eq:winofact}) grows substantially. As shown below, the role of the charginos becomes more and more crucial as the neutralino mass is increased.

\begin{figure}[t!]
	\begin{center}
	\includegraphics[width=\linewidth]{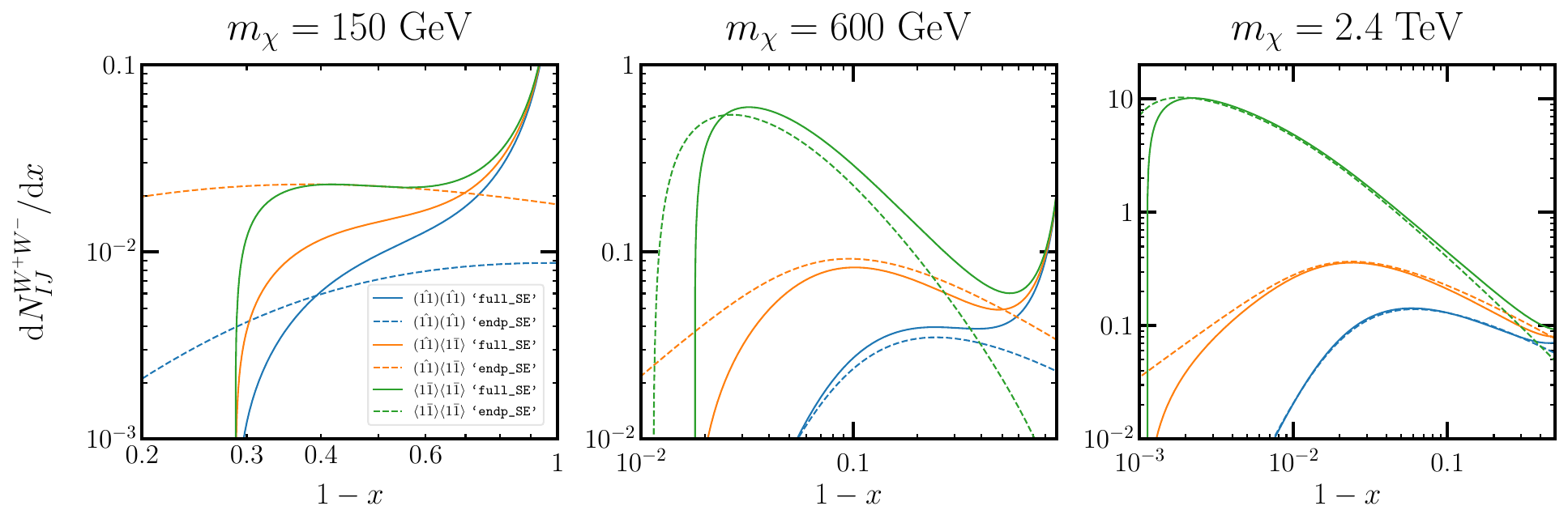} 
	\caption{\label{fig:factthm} 
	Comparison of the normalized matrix elements $\dif{}N_{IJ}^{W^+W^-}/\dif x$ in the pure wino model, as obtained using our \full{} calculation (solid lines), with the \endp{} calculations of Refs.~\cite{Beneke:2019vhz,Beneke:2022eci} (dashed lines), as functions of $(1-x)$ for three different values of $\Mdm$. Each color corresponds to the contribution of a single matrix element with indices $IJ$ as introduced in Sec.~\ref{sec:somm}.  }
	\end{center}
\end{figure}

Our \full{}  calculation exhibits an interesting phenomenon at gamma-ray energies that are somewhat smaller than the  threshold energy of the $W^+W^-\gamma$ final state, $\Eg=\Mdm-m_W^2/\Mdm$. Specifically, unlike the relatively ``soft cutoff'' behavior that is observed in the \fix{} computation, the \full{} prediction exhibits a subtle additional enhancement followed by a sharper decline near the threshold. This additional enhancement becomes more pronounced as the neutralino mass is increased, as apparent from the two $\Mdm$ scenarios depicted in Fig.~\ref{fig:doubleplot}.
In order to better understand this behavior, we introduce the ``normalized'' matrix elements for the $W^+W^-\gamma$ channel, which is the only annihilation channel with non-vanishing diagonal and off-diagonal matrix elements in the wino/higgsino scenarios,  
\begin{equation}
\frac{\dif{}N_{IJ}^{W^+W^-}}{\dif x}\equiv\frac{\Mdm}{\tsv^{W^+W^-}_{IJ}}\frac{\dif{}}{\dif\Eg}\tsv_{IJ}^{W^+W^-\gamma}\ ,
\label{eq:normannmat}
\end{equation}
where $x\equiv\Eg/\Mdm$. 
These matrix elements are dimensionless and approach the LO fragmentation function of  Eq.~\eqref{eq:fsrlo} asymptotically for sufficiently large $\Mdm$ and small enough~$x$. We plot them in Fig.~\ref{fig:factthm}. 
%
We observe that in the regime $(1-x)\ll1$ and for scenarios with heavy neutralinos, such as the considered case of $\Mdm=2.4$~TeV, the diagonal matrix element with $I=J=\chbch 11$ is much larger than the corresponding matrix elements for all other $IJ$ combinations. In this case the three relevant Sommerfeld matrix elements $S_{IJ}$ listed in Tab.~\ref{tab:snumbers} are comparable in order of magnitude. We see that, when they are combined with the relevant annihilation matrix elements, the last term of Eq.~\eqref{eq:winofact} in this scenario gives the dominant contribution in the $(1-x)\ll1$ regime. The endpoint spectrum is thus dominated by virtual chargino annihilations. This is a consequence of the increasing influence of Sudakov logarithms for $(1-x)\ll1$, which are particularly large for the terms of Eq.~\eqref{eq:winofact} involving charginos.

Besides the Sommerfeld effect, which is the focus of this work, a different kind of resummation is necessary at
\emph{the endpoint region} of the spectrum. Indeed, for very heavy neutralinos and $\Eg\to\Mdm$ our \full{} computation suffers from large logarithmic enhancements that need to be incorporated (resummed) into the prediction. 
Such a resummation has already been completed for the MSSM in Ref.~\cite{Beneke:2022pij} at the next-to-leading logarithmic (NLL) accuracy, while in Refs.~\cite{Beneke:2018ssm,Beneke:2019vhz} and \cite{Beneke:2019gtg} an NLL-prime (NLL') accuracy\footnote{In contrast to standard NLL computations, NLL' resumations require that all relevant hard, soft and jet functions are obtained at 1-loop, see e.~g.~Ref.~\cite{Becher:2014oda}. }  was achieved for wino- and higgsino-like DM, respectively (see Ref.~\cite{Vollmann:2022cjc} for a short review). The NLL' calculation exhibits an accuracy of $\mathcal O$(1\%), while the accuracy of the NLL prediction is of order $\mathcal O$(10\%). Similar calculations for wino-like DM can be found in Ref.~\cite{Baumgart:2017nsr} at leading-logarithmic (LL) accuracy and Ref.~\cite{Baumgart:2018yed}~(NLL), and for higgsino-like DM in Ref.~\cite{Baumgart:2015bpa}~(LL'), where all anomalous dimensions are given at 1-loop accuracy.  

Indeed, we also verified that our results are consistent with the endpoint resummed calculation of Refs.~\cite{Beneke:2019vhz,Beneke:2019gtg,Beneke:2022eci}, to which we refer as \endp{} in the following, within their kinematic regimes of validity, namely, $\Mdm-\Eg\sim\ord(m_W)$. For these comparisons an additional step is required. While in the \endp{} calculations, each annihilation-matrix element $\left[\dif{}\tsv/\dif\Eg\right]_{IJ}$ incorporates terms resummed to all orders in the electroweak coupling, annihilation matrix elements in the \full{} computation are computed at LO. Thus, for a meaningful comparison between the two calculations, a fixed-order expansion of each annihilation matrix element of the \endp{} calculation must be performed. The LO term of the fixed-order expansion of the $IJ$-th annihilation matrix element in the \endp{} calculation should then agree, up to power corrections of $\ord(m_W^2/\Mdm^2)$ that are not included in our calculation, with the respective matrix element of the \full{} computation. Indeed, we find that the annihilation matrix elements associated with the $\cha1\acha1\to\gamma+f\bar f$, $\cha1\acha1\to\gamma+Zh$ and $\cha1\acha1\to\gamma+W^+W^-$ processes 
\footnote{
These are the only calculations of chargino-antichargino annihilation cross sections into three-body final states existing in the literature. 
}
are exactly reproduced by the expanded results of Refs.~\cite{Beneke:2019vhz,Beneke:2022eci} at the leading order in the power expansion assumed there, and provided that $1-x\ll 1$. Note that, at LO, the $\gamma+f\bar f$ and $\gamma+Zh$ final states can only occur when the initial state is composed of charginos, i.e.\ these channels only contribute to the diagonal $I=J=\chbch 11$ annihilation matrix element. In contrast, $\gamma+W^+W^-$ final states can also result from neutralino annihilation, which receives contributions also from the $IJ=\nn{11}\nn{11}$ and  $IJ=\nn{11}\chbch 11$ combinations.

The agreement between our calculated annihilation matrix elements and those extracted from Refs.~\cite{Beneke:2019vhz,Beneke:2022eci} is very good for masses  above the TeV~scale, but poor for lighter neutralinos. This is to be expected, as the \endp{}  results  are valid only up to power corrections of $\mathcal O(m_W^2/\Mdm^2)$ which are not negligible for smaller neutralino masses such as $\Mdm=150$~GeV. Our calculations, instead, provide the most accurate and reliable picture for the gamma-ray spectrum from neutralino annihilations in that mass range.

In addition to the comparison with the \endp{} calculation, we compared our results for the $I=J=\nn{11}$ annihilation matrix element analytically with the fixed-order neutralino annihilation cross section into a $W^+W^-\gamma$ final state in the pure wino and higgsino limits that has been provided in Ref.~\cite{Bergstrom:2005ss}. We found exact agreement. 
\section{Conclusions}
\label{sec:conclusions}
In this work, we have presented the first calculation of the continuum gamma-ray spectra resulting from  neutralino annihilation including  the Sommerfeld effect in the MSSM.  The main novelty of our work is the systematic inclusion of all combinations of chargino-antichargino and neutralino pair annihilation processes into three-body final states that play a role in the calculation. The impact of the Sommerfeld effect is generically very strong due to the highly non-relativistic nature of the DM particles  in  the considered scenarios. For the sake of concreteness we focused our numerical discussion on the pure wino and pure higgsino limits of the MSSM. In the neutralino mass range  of about 100~GeV~to~1~TeV, we find qualitative differences compared to calculations of continuum spectra including only final-state radiation that are traditionally employed in gamma-ray searches for WIMP DM. 
For neutralinos heavier than about a TeV, the endpoint of the continuum requires the resummation of  large Sudakov logarithms from soft/collinear electroweak radiation. Our work  fills the gap between previous calculations focusing on the endpoint regime and separate ones for the low-energy photon regime where the widely used final-state radiation approximation is appropriate.

To ensure the correctness of our calculations we performed several stringent consistency checks. We verified that our results agree with older ones in the appropriate limits, we checked gauge invariance, unitary safety, and we re-derived the non-relativistic potential for the $s$-wave annihilation of neutralinos. We also showed that our results are consistent with existing results employing the collinear approximation.

On the technical side, our results will allow for a reassessment of the impact of internal bremsstrahlung  by combining it with the Sommerfeld enhancement effect. This will pave the way for robust global fits, especially in reduced-parameter MSSM scenarios. 
Most importantly, though, our calculation will open the door to detailed phenomenological studies of the indirect detection of neutralinos using gamma-ray observations from \emph{both} satellite and Cherenkov telescopes. In the light of improved energy resolutions and sensitivities of current and next-generation gamma-ray telescopes this is a very timely achievement.

\section*{Acknowledgements}
We are grateful for valuable discussions to Denys Malyshev and Andrea Santangelo.  We would also like to extend our gratitude to Torsten Bringmann for his helpful feedback. We acknowledge support by the state of Baden-W\"urttemberg through bwHPC and the German Research Foundation (DFG) through grant no.~INST 39/963-1 FUGG (bwForCluster NEMO).

\appendix
\section{Conventions}
\label{app:conventions}
\subsection{Neutralino and chargino mixing matrices}
All our computations within the MSSM employed the model file \texttt{MSSM.mod} that is contained in {\tt FeynArts}. 
Detailed information about the conventions used in this model file is given in  Refs.~\cite{Hahn:2001rv,Haber:1984rc,Gunion:1984yn,Gunion:1986nh}. Here, we explicitly denote the elements of the neutralino ($\mathbb M_N$) and chargino ($\mathbb M_C$) mixing matrices, since these are of immediate relevance for the computations that we present in this work. These are given by 
\begin{equation}
\mathbb M_N = 
\begin{pmatrix}
M_1 & 0  & -m_Zs_W\cos\beta & m_Zs_W\sin\beta \\
0 & M_2  & m_Zc_W\cos\beta & -m_Zc_W\sin\beta \\
-m_Zs_W\cos\beta & m_Zc_W\cos\beta & 0 & -\mu \\
m_Zs_W\sin\beta & -m_Zc_W\sin\beta & -\mu & 0 
\end{pmatrix} \ ,
\label{eq:mssmneumatrix}
\end{equation}
and
\begin{equation}
\mathbb M_C = 
\begin{pmatrix}
M_2 & \sqrt2\, m_W\sin\beta \\
\sqrt2\, m_W\cos\beta & \mu
\end{pmatrix}\ ,
\label{eq:mssmchamatrix}
\end{equation}
respectively.

The \emph{unitary} mixing matrices $\tilde N$, $\tilde U$ and $\tilde V$ are  defined by the following conditions: 
\begin{equation}
\label{eq:mixingmatrices}
\tilde N^*\mathbb M_N\tilde N^{-1} =\diag(m_{\neu1},m_{\neu2},m_{\neu3},m_{\neu4})\ , \ \tilde U^*\mathbb M_C\tilde V^{-1}=\diag(m_{\chapm1},m_{\chapm2})\,,
\end{equation}
where $m_{\neu1}<m_{\neu2}<m_{\neu3}<m_{\neu4}$ and $m_{\neu1}<m_{\chapm1}<m_{\chapm2}$.

\subsection{Static potential} 
The potential matrix $V_{IJ}(r)$ for the MSSM given by Eq.~\eqref{eq:mssmpot} can be decomposed as
\beq
V_{IJ}(r)= V_{IJ}^\textrm{diag}(r)+V_{IJ}^\textrm{off}(r)\,, 
\eeq
where
\begin{align*}
V^{\textrm{diag}}_{IJ}(r) &=- \frac{\hv{\gamma}_{IJ}}r - \hv{Z}_{IJ}\frac{\e^{-m_Z r}}r -\hv{h}_{IJ}\frac{\e^{-m_h r}}r -\hv{H^0}_{IJ}\frac{\e^{-m_{H^0} r}}r-\hv{A^0}_{IJ}\frac{\e^{-m_{A^0} r}}r\,,
\\
V^{\textrm{off}}_{IJ}(r) &=- \hv{W}_{IJ}\frac{\e^{-m_W r}}r - \hv{H^+}_{IJ}\frac{\e^{-m_{H^+} r}}r\, .
\end{align*}

The coupling matrices $\hv{B}_{IJ}$ for each mediating boson $B$ are block-diagonal for the neutral mediators $B^0=\gamma,Z,h,H^0,A^0$ and off-diagonal for the charged bosons $B^+=W^+, H^+$, 
\beq
\hv{B^0}_{IJ}=
\begin{pmatrix} 
\hv{B^0}_{\nn{ij},\nn{kl}} & \mathbb O_{10\times4} \\ 
\mathbb O_{4\times10}  & \hv{B^0}_{\chbch xy,\chbch zw} 
\end{pmatrix}
\quad ,\quad\hv{B^+}_{IJ}=
\begin{pmatrix} 
\mathbb O_{10\times10} & \hv{B^+}_{\nn{ij},\chbch zw} \\ 
\hv{B^+}_{\chbch xy,\nn{kl}} & \mathbb O_{4\times4} 
\end{pmatrix}\,.
\eeq
In Eqs.~\eqref{eq:nnZnnpot}-\eqref{eq:nnHccpot} the coupling matrices have been expressed in terms of the coefficients that are displayed below. In particular, the coefficients associated with massive vector boson interactions are given by
\begin{equation}
v_{ij}^{Z\,(0)} = \frac1{2\,c_W}\im\left(N_{i3}N_{j3}^*-N_{i4}N_{j4}^*\right) \quad ,\quad a_{ij}^{Z\,(0)} = \frac1{2\,c_W}\re\left(N_{i3}N_{j3}^*-N_{i4}N_{j4}^*\right) \ ,
\label{eq:vaZ0}
\end{equation}
\begin{equation}
v_{xy}^Z = - \frac1{4c_W}\left(\tilde U_{x1}\tilde U_{y1}^*+\tilde V_{x1}^*\tilde V_{y1}+2(c_W^2-s_W^2)\delta_{xy}\right)
\quad ,\quad
a_{xy}^Z = -\frac1{4c_W}\left(\tilde U_{x1}\tilde U_{y1}^*-\tilde V_{x1}^*\tilde V_{y1}\right) \ ,
\label{eq:vaZ}
\end{equation}
\begin{eqnarray}
v_{ix}^W  &=& \frac12\left[\tilde N_{i2}^*\tilde U_{x1}+\tilde N_{i2}\tilde V_{x1}^*+\frac1{\sqrt2}\left(\tilde N_{i3}^*\tilde U_{x2}-\tilde N_{i4}\tilde V_{x2}^*\right)\right] \ ,\\
a_{ix}^W &=& \frac12\left[\tilde N_{i2}^*\tilde U_{x1}-\tilde N_{i2}\tilde V_{x1}^*+\frac1{\sqrt2}\left(\tilde N_{i3}^*\tilde U_{x2}+\tilde N_{i4}\tilde V_{x2}^*\right)\right]\ ,
\label{eq:vaW}
\end{eqnarray}
where $c_W=\cos\theta_W$ and $s_W=\sin\theta_W$, and the $N_{ij}$, $\tilde U_{ij}$ and $\tilde V_{ij}$ denote elements of the unitary mixing matrices defined in Eq.~\eqref{eq:mixingmatrices}.  
The coefficients that are associated with charged scalar bosons read
\begin{align}
s_{ix}^{G_W}  =  -\frac1{2\,c_W}{}&{}\left[s_\beta\frac{s_W\tilde N_{i1}^*+c_W\tilde N_{i2}^*}{\sqrt2}\tilde V_{x2}^*+c_{\beta}\frac{s_W\tilde N_{i1}+c_W\tilde N_{i2}}{\sqrt2}\tilde U_{x2}\right.\nonumber\\
{}&{}\left.-c_\beta c_W\tilde N_{i3}\tilde U_{x1}+s_\beta c_W\tilde N_{i4}^*\tilde V_{x1}^*\right] \ ,
\label{eq:sGW}
\\
s_{ix}^{H^+} =  -\frac1{2\,c_W}{}&{}\left[c_\beta\frac{s_W\tilde N_{i1}^*+c_W\tilde N_{i2}^*}{\sqrt2}\tilde V_{x2}^*-s_{\beta}\frac{s_W\tilde N_{i1}+c_W\tilde N_{i2}}{\sqrt2}\tilde U_{x2}\right.\nonumber\\
{}&{}\left.+s_\beta c_W\tilde N_{i3}\tilde U_{x1}+c_\beta c_W\tilde N_{i4}^*\tilde V_{x1}^*\right]\ ,
\end{align}
where $c_\beta\equiv\cos\beta$ and $s_\beta\equiv\sin\beta$. Finally, the coefficients for the neutral scalar bosons are given by  
\begin{eqnarray}
s_{ij}^{G_Z\,(0)}  &=& \frac{i}{4c_W}\left\{[(c_\beta N_{i3}+s_\beta N_{i4})(s_W N_{j1}-c_W N_{j2})+(i\leftrightarrow j)]-\textrm{c.~c.}\right\}\ ,
\label{eq:sGZ0}
\\
s_{ij}^{A_0\,(0)}  &=& -\frac{i}{4c_W}\left\{[(s_\beta N_{i3}-c_\beta N_{i4})(s_W N_{j1}-c_W N_{j2})+(i\leftrightarrow j)]-\textrm{c.~c.}\right\}\ ,\\
s_{ij}^{H\,(0)}  &=& \frac1{4c_W}\left\{[(c_\alpha N_{i3}-s_\alpha N_{i4})(s_W N_{j1}-c_W N_{j2})+(i\leftrightarrow j)]+\textrm{c.~c.}\right\}\ ,\\
s_{ij}^{h\,(0)}  &=& -\frac1{4c_W}\left\{[(s_\alpha N_{i3}+c_\alpha N_{i4})(s_W N_{j1}-c_W N_{j2})+(i\leftrightarrow j)]-\textrm{c.~c.}\right\}\ ,
\end{eqnarray}
\begin{eqnarray}
s_{xy}^{G_Z}  &=& -\frac{i}{2\sqrt2}\left[c_\beta(U_{x2}V_{y1}-U_{y2}^*V_{x1}^*)-s_\beta(U_{x1}V_{y2}-U_{y1}^*V_{x2}^*)\right]\ ,
\label{eq:sGZ}
\\
s_{xy}^{A_0}  &=& \frac{i}{2\sqrt2}\left[s_\beta(U_{x2}V_{y1}-U_{y2}^*V_{x1}^*)+c_\beta(U_{x1}V_{y2}-U_{y1}^*V_{x2}^*)\right]\ ,\\
s_{ij}^{H}  &=& -\frac1{2\sqrt2}\left[c_\alpha (U_{x2}V_{y1}+U_{y2}^*V_{x1}^*)+s_\alpha (U_{x1}V_{y2}+U_{y1}^*V_{x2}^*)\right]\ ,\\
s_{ij}^{h}  &=& \frac1{2\sqrt2}\left[s_\alpha (U_{x2}V_{y1}+U_{y2}^*V_{x1}^*)-c_\alpha (U_{x1}V_{y2}+U_{y1}^*V_{x2}^*)\right]\,,
\end{eqnarray}
where $s_\alpha\equiv\sin\tilde\alpha$, $c_\alpha\equiv\cos\tilde\alpha$ and $\tilde\alpha$ is the mixing angle of the CP-even Higgs doublet.

\section{Parameterization of the three-body phase space}
\label{app:phase-space}
In this work, we consider neutralino annihilation processes into three-particle final states of the type $\gamma+X^{(2)}$.  The four-momenta of the particles are denoted by $k_i$ with $(i=1,\ldots 5)$ such that 
\begin{equation}
\chi_0(k_1) + \chi_0(k_2) \to X_1^{(2)}(k_3) +X_2^{(2)}(k_4) +  \gamma(k_5)\,, 
\end{equation} 
with the $X^{(2)}_{1,2}$ denoting the two final-state particles resulting from the $X^{(2)}$ system. 

The non-relativistic initial-state particles are assumed to be at rest, i.e.\ \(k_1=(\Mdm,\vec0)^T\) and \(k_2=(\Mdm,\vec0)^T\).  
Assuming that the photon propagates along the $z$ axis we find 
\[k_5^0=||\vec k_5||=m_\chi x=m_\chi-\frac{s_{34}}{4m_\chi}\,, \quad \textrm{with }\ s_{34}=m_X^2=(k_3+k_4)^2 =4m^2_\chi(1-x)\ ,\]
where $m_X$ is the invariant mass of the subsystem $X^{(2)}$ of the final state in a $2\to3$ scattering process of type $\neu1\neu1\to\gamma+X^{(2)}$. We  choose a reference system where $\vec k_3$ and $\vec k_4$ are lying in the $yz$ plane ($k_3^x=k_4^x=0$). The components  of $k_3^\mu$ are given by
    \begin{align*}
    k_3^0={}&{}\frac{m_\chi}{2z}\left[(1+z)\left(z-\frac{m_4^2-m_3^2}{4m_\chi^2}\right)-(1-z)\sqrt{z-\left(\frac{m_4-m_3}{2m_\chi}\right)^2}\sqrt{z-\left(\frac{m_4+m_3}{2m_\chi}\right)^2}\cos\theta_3^*\right]\,,\\
     k_3^y={}&{}-k_4^y=\frac{m_\chi}{\sqrt{z}}\,\sqrt{z-\left(\frac{m_4-m_3}{2m_\chi}\right)^2}\sqrt{z-\left(\frac{m_4+m_3}{2m_\chi}\right)^2}\sin\theta^*_3\,,\\
     k_3^z={}&{}-\frac{m_\chi}{2z}\left[(1-z)\left(z-\frac{m_4^2-m_3^2}{4m_\chi^2}\right)-(1+z)\sqrt{z-\left(\frac{m_4-m_3}{2m_\chi}\right)^2}\sqrt{z-\left(\frac{m_4+m_3}{2m_\chi}\right)^2}\cos\theta_3^*\right]\,, 
    \end{align*}
where $z=1-x$ with $x=\Eg/\Mdm$; $\theta_3^*$ ($\theta_4^*$) is the relative angle between $k_3$ ($k_4$) and $k_5$ in the rest frame of $X^{(2)}$. The components of $k_4$ are obtained by interchanging $m_3$, $\theta_3^*$ with $m_4$ and $\theta_4^*$ 
in the expressions for $k_3$ above. The  three-particle phase space can thus be fully parametrized by the variables $x$ and $\theta_3^*$. 

The annihilation matrix elements of Eq.~\eqref{eq:annmat} in this parametrization are given by 
    \begin{equation}
    \label{eq:2to3annmatrix}
\left[\frac{{\rm d}\tsv{}}{{\rm d}\Eg}\right]_{IJ}=\frac1{(\sqrt2)^{\id(I)+\id(J)}}\sum_{X^{(2)}}\frac{x\lambda\left[4\,m_\chi^2(1-x),m_3^2,m_4^2\right]}{(16\,\pi)^3m_\chi^3(1-x)}
\int_{-1}^1 {\rm d}(\cos\theta_3^*)
\amp^{(l,s)=(0,0)\,*}_{J\to\gamma+X^{(2)}}\amp^{(l,s)=(0,0)}_{I\to\gamma+X^{(2)}}\,, 
\end{equation}
where $\lambda[a,b,c]=\sqrt{a^2+b^2+c^2-2ab-2ac-2bc}$. In our generic notation the summation over all possible spin and helicity configurations of the final-state particles is implied in the summation over $X^{(2)}$.

\bibliographystyle{JHEP}


\bibliography{biblio}
\end{document}